\documentclass[10pt,pra,twocolumn,showpacs,superscriptaddress,floatfix]{revtex4-2}
\usepackage{graphicx}
\usepackage{amssymb}
\usepackage{amsmath}
\usepackage{amsthm}
\usepackage{bm}
\usepackage{physics}
\usepackage{color,xcolor}
\usepackage{subfigure}
\usepackage{algpseudocode}
\usepackage{algorithm}
\usepackage{algorithmicx}
\usepackage{lineno}
\usepackage{soul}
\usepackage{mathtools} 
\usepackage{mathrsfs}
\usepackage{graphicx}
\usepackage{dcolumn}
\usepackage{textcomp,mathcomp}
\usepackage{enumerate}

\usepackage{lipsum} 
\usepackage[colorlinks,
linkcolor=blue,      
anchorcolor=blue, 
citecolor=purple]{hyperref}

\begin{document}

\preprint{AAPM/123-QED}

\title{Preserving Heisenberg-Limited Metrological Information during Storage via Correlated-Noise Correction}

\author{Hang Xu}
\affiliation{State Key Laboratory of Photonics and Communications, Institute for Quantum Sensing and Information Processing, Shanghai Jiao Tong University, Shanghai 200240, People’s Republic of China}%


\author{Xue-Ke Song}
\affiliation{School of Physics, Anhui University, Hefei 230601,  , People’s Republic of China}

\author{Jingzheng Huang}%
\affiliation{State Key Laboratory of Photonics and Communications, Institute for Quantum Sensing and Information Processing, Shanghai Jiao Tong University, Shanghai 200240, People’s Republic of China}%
\affiliation{Hefei National Laboratory, Hefei, 230088, People’s Republic of China}
\affiliation{Shanghai Research Center for Quantum Sciences, Shanghai, 201315, People’s Republic of China}

\author{Tailong Xiao}
\email{tailong\_shaw@sjtu.edu.cn}
\affiliation{State Key Laboratory of Photonics and Communications, Institute for Quantum Sensing and Information Processing, Shanghai Jiao Tong University, Shanghai 200240, People’s Republic of China}%
\affiliation{Hefei National Laboratory, Hefei, 230088, People’s Republic of China}
\affiliation{Shanghai Research Center for Quantum Sciences, Shanghai, 201315, People’s Republic of China}

\author{Guihua Zeng}
\affiliation{State Key Laboratory of Photonics and Communications, Institute for Quantum Sensing and Information Processing, Shanghai Jiao Tong University, Shanghai 200240, People’s Republic of China}%
\affiliation{Hefei National Laboratory, Hefei, 230088, People’s Republic of China}
\affiliation{Shanghai Research Center for Quantum Sciences, Shanghai, 201315, People’s Republic of China}

\date{\today}


\begin{abstract}

Quantum error correction has become an indispensable tool for restoring Heisenberg-limited precision in noisy quantum metrology. Existing protocols, however, almost exclusively focus on correcting noise during the signal-encoding stage and implicitly assume that the probe is measured immediately after sensing. In many quantum information processing tasks, the encoded probe must instead be stored before subsequent quantum operations, during which environmental noise can significantly degrade the accumulated metrological information.
Here, we propose a correlated-noise correction (CNC) protocol for protecting quantum probes during the storage stage. By correlating probe errors with auxiliary qubits through fixed two-body entangling gates, memory errors are converted into measurable syndromes that are extracted only once after storage. We show that the protocol naturally extends from single-qubit to multi-qubit probes and protects the stored quantum Fisher information against dephasing, bit-flip, and amplitude-damping noise. Furthermore, we demonstrate that preserving the quantum Fisher information does not necessarily require restoring the entire quantum state when the probe is measured immediately after storage, whereas full state recovery becomes essential for subsequent rounds of quantum signal processing.
Our results establish correlated-noise correction as a practical framework for protecting metrological information during quantum memory and provide a useful building block for sensing-enabled quantum information processing.
\end{abstract}


\maketitle

\section{INTRODUCTION}
Quantum sensing \cite{sensing1,sensing2,sensing3,sensing4} leverages coherence and entanglement to enable precision measurements at the Heisenberg limit \cite{HL1}, whereas classical strategies can only achieve the standard quantum limit. Although it has been demonstrated that the Heisenberg limit can be surpassed in scenarios involving indefinite order causality \cite{ico_cvsensing1,ico_cvsensing2}, non-Hermitian systems \cite{nonh1,nonh2,nonh3}, and phase transitions \cite{HL1,phase1,phase2}, these breakthroughs essentially stem from a redefinition of resources. 
For example, the super-Heisenberg limit based on indefinite order causality actually stems from an infinite-dimensional parameter encoding space, while the precision improvement in non-Hermitian systems arises from probabilistic post-selection.
Furthermore, practical quantum devices are inevitably subject to environmental noise, making it difficult to sustain the quantum advantage in sensing precision \cite{noisysensing0,noisysensing1,noisysensing2,noisysensing3,noisysensing4}. 
The No-Go theorem \cite{nogo} states that achieving the Heisenberg limit in noisy scenarios is impossible through unitary control alone.

Quantum control \cite{control1,control2,control3,control4}, error mitigation \cite{qem1,qem2,qem3}, and dynamic decoupling \cite{dd1,dd2} can reduce or eliminate the effects of noise. 
Quantum control can avoid noise-sensitive regions by guiding the path of unknown parameter encoding. However, it remains fundamentally within the realm of unitary control and is constrained by the No-Go theorem. 
Error mitigation relies on the probability of successful post-selection; if all quantum resources are considered, it cannot guarantee the restoration of the Heisenberg limit. 
Dynamic decoupling is only applicable to non-Markovian noise scenarios; for non-Markovian noise processes, decoupling between noise and the unitary evolution cannot be guaranteed. 

Quantum error correction (QEC) \cite{qec1,qec2,qec3,qec4,qec5,qec6,qec7,correction1,correction2} provides a general framework for suppressing noise and is one of the fundamental techniques for realizing large-scale fault-tolerant quantum information processing. In quantum metrology, QEC employs noise-free auxiliary qubits and syndrome measurements to continuously remove errors while preserving signal accumulation, thereby recovering Heisenberg-limited precision under suitable conditions. Since the signal Hamiltonian and environmental noise act simultaneously during the sensing process, conventional QEC generally requires the noise operators to be orthogonal to the signal Hamiltonian \cite{qec5}; otherwise, syndrome extraction inevitably removes part of the accumulated signal together with the noise. More recently, this restriction has been relaxed by extending quantum metrology to the indefinite causal order regime, where signal reversal enables even parallel noise to be corrected \cite{qec7}.

Despite these advances, existing QEC protocols implicitly assume that the sensing task terminates immediately after signal encoding. Once the desired parameter has been accumulated, the probe is directly measured, so all errors are assumed to occur only during the sensing stage. This assumption is well suited for conventional quantum metrology, where estimating the unknown parameter is the final objective.

However, this picture changes fundamentally when quantum sensing is embedded into a larger quantum-information-processing workflow. In many emerging applications, sensing is no longer the final step but instead serves as one computational primitive among many. Rather than being measured immediately, the quantum probe carrying the encoded parameter must be preserved until subsequent quantum operations are completed. 
During this storage interval, no additional signal is accumulated, yet environmental noise continuously degrades the encoded metrological information. 
Such scenarios naturally arise in quantum machine learning \cite{nisq1,qml1,qml2,qml3,qml4,qml5}, where gradients encoded by quantum sensing circuits may need to be accumulated over many optimization iterations before parameter updates are performed; in variational quantum algorithms \cite{vqa1,vqa2,vqa3}, where intermediate quantum states must be preserved while expectation values from different circuits are jointly processed; and in distributed quantum sensing \cite{dqs1,dqs2} and quantum networks\cite{qn1,qn2}, where probe states carrying local sensing information must be stored while waiting for synchronization, classical communication, or subsequent quantum operations.

Consequently, the dominant source of precision loss may no longer originate from the sensing stage itself but from the subsequent storage stage. Existing QEC protocols are not designed for this situation because signal accumulation has already been completed and the objective is no longer to protect the sensing dynamics, but rather to preserve the metrological information that has already been encoded. Unlike QEC during sensing, memory protection does not necessarily require restoring the entire quantum state. As we will show below, if the probe is measured immediately after storage, preserving the quantum Fisher information can be sufficient even when the stored quantum state differs from the ideal one. On the other hand, if the stored probe is intended for subsequent signal accumulation or quantum-information processing, faithful recovery of the probe state becomes essential. These observations motivate the development of a dedicated error-correction framework for the quantum memory stage.

In this work, we propose a correlated-noise correction (CNC) protocol for protecting quantum probes during the storage stage after signal encoding. The central idea is to correlate memory errors of the probe with auxiliary qubits through fixed two-body entangling gates, thereby converting probe errors into measurable syndromes that are extracted only once at the end of the storage process. We first establish a general CNC framework and show that it can be naturally extended from single-qubit to multi-qubit systems. We then demonstrate its effectiveness for preserving the quantum Fisher information under dephasing, bit-flip, and amplitude-damping noise. More importantly, we show that protecting metrological information during storage is fundamentally different from protecting the sensing process itself. If the probe is measured immediately after storage, preserving the quantum Fisher information does not necessarily require restoring the entire quantum state. In contrast, when the stored probe is reused for subsequent signal accumulation or quantum-information processing, faithful recovery of the probe state becomes essential. These results establish CNC as a practical error-correction framework for protecting metrological information during quantum memory and provide a useful building block for sensing-enabled quantum information processing.

\section{Noisy Quantum Sensing}

We consider the estimation of an unknown parameter $\omega$ encoded through a noiseless sensing channel
\begin{equation}
{\cal E}(t_s)=e^{-iH_st_s},
\end{equation}
where the sensing Hamiltonian is
\begin{equation}
H_s=\omega G.
\end{equation}
After a sensing duration $t_s$, the precision of an unbiased estimator $\hat\omega$ is bounded by the Cramér-Rao inequality \cite{crb1,crb2}
\begin{equation}
{\rm var}(\hat\omega)
\ge
{\cal M}^{-1}F_C^{-1}
\ge
{\cal M}^{-1}F_Q^{-1},
\end{equation}
where ${\cal M}$ is the number of measurement shots, while $F_C$ and $F_Q$ denote the classical and quantum Fisher information, respectively. Throughout this work, we adopt the QFI as the figure of merit for sensing precision, which is evaluated through
\begin{equation}
F_Q=
\frac{8\left[
1-
F(\rho_\omega,\rho_{\omega+d\omega})
\right]}
{d\omega^2},
\end{equation}
with $F(\rho_\omega,\rho_{\omega+d\omega})$ the quantum-state fidelity.

For a single-qubit probe prepared in $|+\rangle$ with
\begin{equation}
H_s=\omega\sigma_z,
\end{equation}
the noiseless sensing process yields
\begin{equation}
F_Q=4t_s^2.
\end{equation}
For an $N$-qubit GHZ probe with collective sensing Hamiltonian
\begin{equation}
H_s
=
\omega
\sum_{i=1}^{N}
\sigma_z^{(i)},
\end{equation}
the QFI becomes
\begin{equation}
F_Q=4N^2t_s^2,
\end{equation}
corresponding to Heisenberg-limited scaling with respect to both sensing time and probe number.

Unlike conventional quantum metrology, we assume that the sensing stage has already been protected from noise by existing QEC techniques, or equivalently that the sensing duration is much shorter than the subsequent storage time. Consequently, the dominant source of precision degradation originates from the storage stage rather than the sensing stage.

After signal encoding, the probe enters a memory channel $\mathscr{E}(t_m)$, where no signal Hamiltonian is present. During the storage time $t_m$, environmental noise continues to act on the probe, and the memory channel may additionally contain a static drift Hamiltonian generated by uncontrolled background fields. The final probe state is therefore
\begin{equation}
\rho_f
=
\mathscr{E}(t_m)
\circ
{\cal E}(t_s)
(\rho_0).
\end{equation}

The objective of this work is to preserve the metrological information encoded during the sensing stage throughout the subsequent storage process, thereby enabling Heisenberg-limited precision even when the probe cannot be measured immediately after signal accumulation.

\begin{figure}[htbp]
  \centering
  \includegraphics[width=0.9\linewidth]{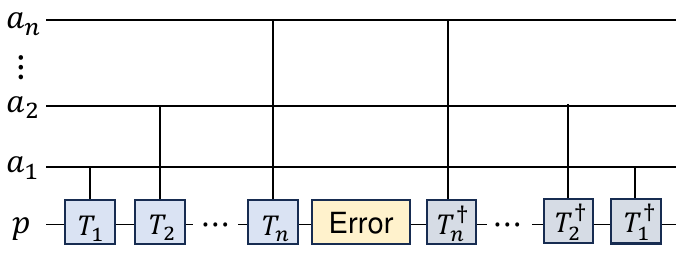}
    \caption{
General framework of the correlated-noise correction (CNC) protocol. Before the memory stage, each auxiliary qubit is entangled with the probe through a two-body gate $T_i$. During the memory process, probe errors are coherently correlated with the corresponding auxiliary qubits. After storage, the inverse gates $T_i^\dagger$ transform the accumulated probe errors into measurable auxiliary syndromes, which are extracted by measuring the auxiliary qubits. The probe itself is never measured during the memory process.
}
  \label{frame}
\end{figure}

\section{Correlated-Noise Correction Protocol}

The central idea of CNC protocol is to transfer memory errors on the probe into correlated errors between the probe and auxiliary qubits. Unlike conventional quantum error correction, no syndrome measurement is performed during the memory process. Instead, the auxiliary qubits coherently record the occurrence of memory errors throughout the storage stage, and all syndrome information is extracted only once after the memory process has finished.

The general structure of the CNC protocol is illustrated in Fig.~\ref{frame}. Consider a memory channel described by $n$ independent noise operators acting on the probe,
\[
\{L_1^{(p)},L_2^{(p)},\cdots,L_n^{(p)}\}.
\]
To monitor these errors, we introduce $n$ auxiliary qubits $\{a_1,a_2,\cdots,a_n\}$, each coupled to the probe through a two-body entangling gate $T_i$. Before the memory channel, the entangling gates $T_i$ sequentially correlate the probe with the corresponding auxiliary qubits. After the memory process, the inverse gates $T_i^\dagger$ are applied in the reverse order, converting the accumulated probe errors into measurable auxiliary syndromes.

To correlate one particular probe error with one auxiliary qubit while leaving all remaining error channels unaffected, the entangling gate $T_i$ is required to satisfy
\begin{equation}
\begin{aligned}
T_iL_j^{(p)}&=L_j^{(p)}T_i,\qquad (j\neq i),\\
T_i^\dagger L_i^{(p)}T_i
&=
P^{(a_i)}\otimes L_i^{(p)},
\end{aligned}
\label{condition1}
\end{equation}
where $P^{(a_i)}$ denotes a Pauli operator acting on the $i$th auxiliary qubit.

The first condition guarantees that the entangling gate only acts on the target error channel, while the second condition transforms the probe error into a correlated probe--auxiliary error. Consequently, after all entangling gates are applied, the memory errors become
\begin{equation}
\begin{aligned}
&T_1^\dagger\cdots T_n^\dagger
L_n^{(p)}\cdots L_1^{(p)}
T_n\cdots T_1 \\
&\qquad =
\left(P^{(a_n)}\otimes L_n^{(p)}\right)
\cdots
\left(P^{(a_1)}\otimes L_1^{(p)}\right).
\end{aligned}
\label{condition2}
\end{equation}
It shows that each independent memory error is mapped onto an independent auxiliary syndrome. Therefore, if every auxiliary qubit is initialized in the state $|+\rangle$, an error on the probe flips the corresponding auxiliary qubit to $|-\rangle$. By measuring the auxiliary qubits in the $X$ basis after the memory stage, the occurrence of every independent memory error can be identified without directly measuring the probe.

If the corresponding probe error is self-adjoint,
\[
\left(L_i^{(p)}\right)^2=\mathbb{I},
\]
the detected error can be completely removed by applying the same operator $L_i^{(p)}$ to the probe. Therefore, the construction of a CNC protocol reduces to finding a set of entangling gates satisfying Eq.~(\ref{condition1}).

It is worth emphasizing that Eq.~(\ref{condition1}) depends only on the structure of the memory error operators and is completely independent of the sensing Hamiltonian. This is fundamentally different from conventional QEC for quantum metrology, where the signal Hamiltonian constrains the admissible correction conditions through the HNLS criterion. Since signal accumulation has already been completed before the storage stage begins, memory protection is decoupled from signal encoding, leading to a much simpler correction strategy. In the following sections, we demonstrate that suitable entangling gates can be systematically constructed for spin systems and naturally generalized from single-qubit probes to multi-qubit quantum memories.

\begin{figure}[htbp]
  \centering
  \includegraphics[width=1\linewidth]{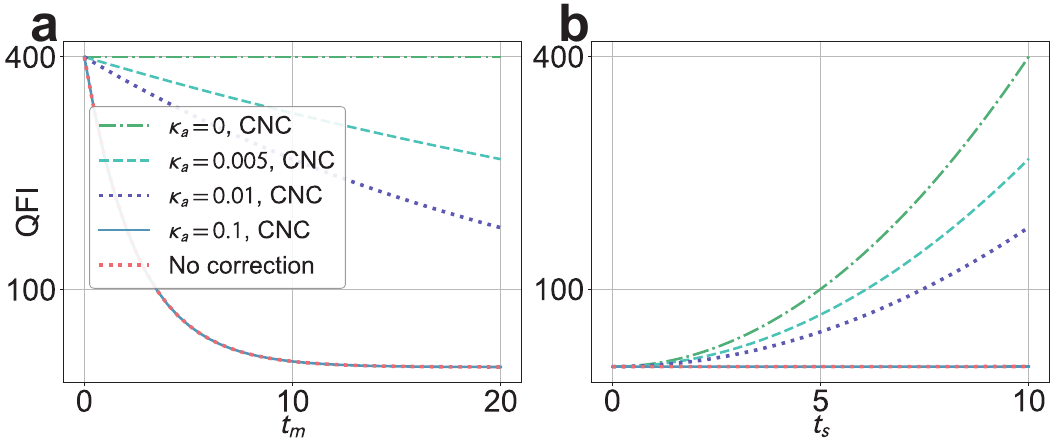}
  \caption{
  Performance of the CNC protocol for a single-qubit probe subject to dephasing noise during the memory stage. (a) QFI as a function of the memory time $t_m$ for a fixed sensing time $t_s=10$, where different curves correspond to different dephasing rates of the auxiliary qubit. (b) QFI as a function of the sensing time $t_s$ for a fixed memory time $t_m=20$. The black dashed line denotes the ideal noiseless result. 
}
  \label{fig2}
\end{figure}

\section{Single-Qubit Probe}

We first illustrate the CNC protocol with a single-qubit probe. The signal is encoded by
\begin{equation}
H_s=\omega\sigma_z^{(p)},
\end{equation}
and the probe is initialized in $|+\rangle_p$. The sensing stage is assumed to be noiseless, either because it has already been protected by conventional QEC or because its duration is much shorter than the subsequent storage time. In the absence of storage noise, the QFI reaches the ideal value $F_Q=4t_s^2$. After sensing, the probe enters a memory channel for a duration $t_m$, during which no signal Hamiltonian acts.

\subsection{Dephasing Noise}

We first consider dephasing noise on the probe, $L_z^{(p)}=\sigma_z^{(p)}$. To detect this error, we introduce one auxiliary qubit initialized in $|+\rangle_a$ and apply the entangling gate
\begin{equation}
T_z=|0\rangle_a\langle0|\otimes I_p
+|1\rangle_a\langle1|\otimes\sigma_x^{(p)} .
\end{equation}
This gate satisfies
\begin{equation}
T_z^\dagger(I_a\otimes\sigma_z^{(p)})T_z = \sigma_z^{(a)}\otimes\sigma_z^{(p)} .
\end{equation}
Thus, a dephasing error on the probe is converted into a correlated error on the auxiliary qubit. Since $\sigma_z^{(a)}|+\rangle_a=|-\rangle_a$, the error can be detected by measuring the auxiliary qubit in the $X$ basis. If the outcome is $|-\rangle_a$, a correction $\sigma_z^{(p)}$ is applied to the probe.

Figure~\ref{fig2} shows the performance of this protocol. In Fig.~\ref{fig2}(a), the sensing time is fixed at $t_s=10$, and the QFI is plotted as a function of the memory time. We also include the case where the auxiliary qubit itself suffers dephasing noise with rate $\kappa_a$. When the auxiliary qubit is noise-free, CNC fully preserves the QFI during storage. As $\kappa_a$ increases, the syndrome information stored in the auxiliary qubit becomes corrupted, leading to a gradual decrease in the QFI. 
Nevertheless, the auxiliary qubit does not need to be loaded with a signal, a noise-free auxiliary qubit can be constructed from a noisy physical qubit using a Shur code. 
Figure~\ref{fig2}(b) shows the QFI as a function of $t_s$ for a fixed memory time $t_m=20$. The noise-free auxiliary case recovers the ideal Heisenberg scaling, while auxiliary noise reduces the achievable precision.

\begin{figure}[htbp]
  \centering
  \includegraphics[width=1\linewidth]{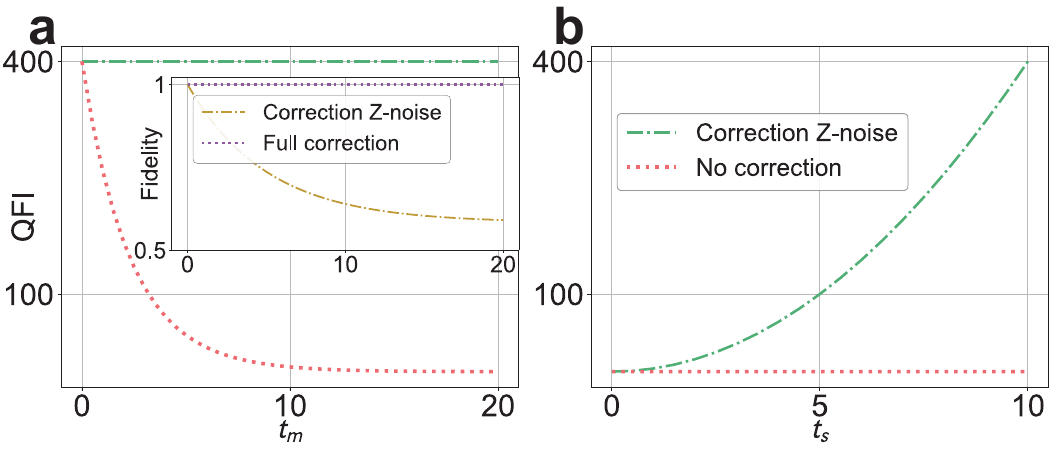}
  \caption{
  Performance of the CNC protocol for a memory channel containing simultaneous bit-flip and dephasing noise. (a) QFI versus the memory time $t_m$ for a fixed sensing time $t_s=10$. The inset shows the fidelity between the stored probe state and the corresponding noiseless state. (b) QFI versus the sensing time $t_s$ for a fixed memory time $t_m=20$. Results obtained by correcting only the dephasing noise are compared with those obtained by correcting both noise channels and with the ideal noiseless case.
}
  \label{fig3}
\end{figure}

\subsection{Bit-Flip and Dephasing Noise}

We next consider a memory channel containing both bit-flip and dephasing noise,
\begin{equation}
L_x^{(p)}=\sigma_x^{(p)},\qquad
L_z^{(p)}=\sigma_z^{(p)} .
\end{equation}
The dephasing component is corrected by the gate $T_z$ introduced above. If bit-flip errors also need to be corrected, a second auxiliary qubit can be introduced with
\begin{equation}
T_x=|0\rangle_a\langle0|\otimes I_p
+|1\rangle_a\langle1|\otimes\sigma_z^{(p)} ,
\end{equation}
which satisfies
\begin{equation}
T_x^\dagger(I_a\otimes\sigma_x^{(p)})T_x = \sigma_z^{(a)}\otimes\sigma_x^{(p)} .
\end{equation}
Thus, bit-flip errors can also be detected by an $X$-basis measurement of the corresponding auxiliary qubit and corrected by applying $\sigma_x^{(p)}$.

A key feature of the memory problem is that the signal has already been encoded before the noisy storage stage begins. Therefore, if the probe is measured immediately after storage, one does not necessarily need to restore the full quantum state. For the present $Z$-signal encoding, correcting only the dephasing component is already sufficient to preserve the encoded parameter information.

This point is demonstrated in Fig.~\ref{fig3}. Figures~\ref{fig3}(a) and (b) show that correcting only dephasing noise gives the same QFI as both the ideal noiseless case and the protocol that corrects both bit-flip and dephasing noise. However, the inset of Fig.~\ref{fig3}(a) shows a different behavior at the state level. When only dephasing noise is corrected, the fidelity between the stored probe and the noiseless probe decreases with memory time, because residual bit-flip errors still modify the state. By contrast, full correction of both noise channels restores the ideal state.

This distinction is important for quantum-information-processing tasks. If the probe is read out immediately after storage, preserving the QFI is sufficient, and partial correction can be optimal. However, if the stored probe must subsequently undergo further signal accumulation, quantum control, or computation, then the probe state itself must be restored. In this case, full correction of both bit-flip and dephasing errors is necessary.

\begin{figure}[htbp]
  \centering
  \includegraphics[width=1\linewidth]{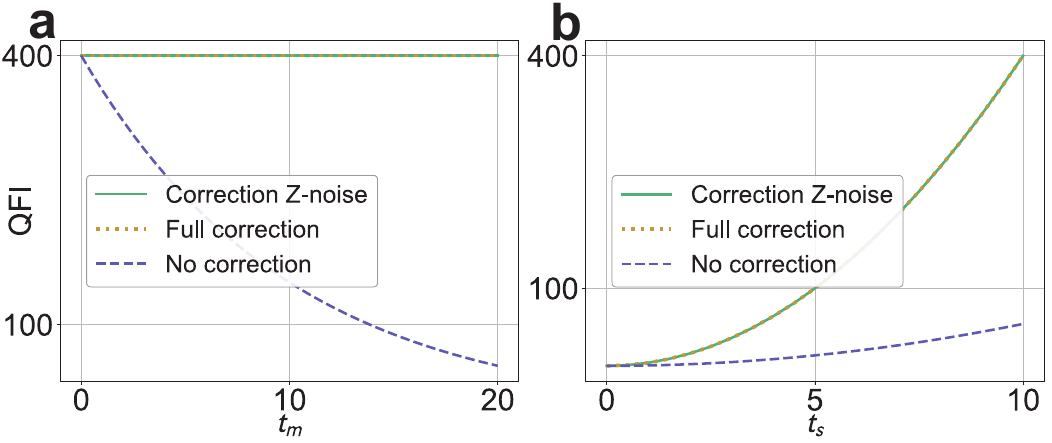}
  \caption{
  Performance of the CNC protocol for amplitude-damping noise during the memory stage. (a) QFI as a function of the memory time $t_m$ for a fixed sensing time $t_s=10$. (b) QFI as a function of the sensing time $t_s$ for a fixed memory time $t_m=20$. The corrected protocol is compared with the uncorrected protocol and the ideal noiseless limit.
}
  \label{fig4}
\end{figure}

\subsection{Amplitude-Damping Noise}

Finally, we consider amplitude-damping noise during the memory stage. The corresponding jump operator is
\begin{equation}
L_-^{(p)}=\sigma_-^{(p)}
=\frac{1}{2}\left(\sigma_x^{(p)}+i\sigma_y^{(p)}\right).
\end{equation}
Unlike Pauli noise, amplitude damping is irreversible and is not directly described by a single self-adjoint error operator. Nevertheless, in the present memory setting, the goal is not necessarily to restore the full probe state, but to preserve the parameter information already encoded before storage.

For the single-qubit probe considered here, the signal is encoded by $\sigma_z^{(p)}$. Therefore, if the probe is measured immediately after the memory stage, it is sufficient to correct the error component that degrades the stored $Z$-signal information. This can be achieved with the same one-auxiliary CNC construction used for dephasing noise. Although amplitude damping also changes the probe state through other effective components, these residual changes do not reduce the QFI associated with the already encoded parameter. Consequently, correcting only the relevant syndrome is enough to recover the ideal QFI.

The numerical results are shown in Fig.~\ref{fig4}. Figure~\ref{fig4}(a) plots the QFI as a function of memory time for fixed $t_s=10$, and Fig.~\ref{fig4}(b) plots the QFI as a function of sensing time for fixed $t_m=20$. Without correction, amplitude damping rapidly reduces the stored sensing precision. In contrast, the one-auxiliary CNC protocol that corrects the relevant $Z$-type syndrome restores the QFI to the ideal value. This shows that even for irreversible amplitude-damping noise, full state recovery is not required if the probe is directly measured after storage.

However, this does not mean that amplitude damping is completely corrected by a single auxiliary qubit. If the stored probe is intended for subsequent signal accumulation, quantum control, or further quantum-information processing, then the residual state distortion induced by amplitude damping must also be removed. In that case, a more complete CNC construction with additional auxiliary qubits is required to monitor and correct all relevant effective error components. Thus, as in the simultaneous bit-flip and dephasing case, amplitude damping highlights the distinction between preserving metrological information and restoring the full quantum state.

\label{sec:single_qubit}

\section{Multi-Qubit Probe}

The CNC protocol can be naturally generalized from a single-qubit probe to arbitrary multi-qubit systems. Consider an $N$-qubit probe subject to independent memory noise,
\begin{equation}
{L_1^{(p)},L_2^{(p)},\cdots,L_m^{(p)}},
\end{equation}
where each noise operator acts locally or collectively on the probe Hilbert space. For every independent error component, one introduces an auxiliary qubit together with an entangling gate $T_i$ satisfying Eq.~(\ref{condition1}),
\begin{equation}
T_i^\dagger L_i^{(p)}T_i=P^{(a_i)}
\otimes
L_i^{(p)}.
\end{equation}
Since different auxiliary qubits are mutually independent, the total transformation is simply the tensor product of the corresponding single-error constructions,
\begin{equation}
T=T_1T_2\cdots T_m .
\end{equation}
Consequently, each independent noise process is mapped onto a distinct auxiliary syndrome and can be detected by measuring the corresponding auxiliary qubit after the memory stage. Therefore, the CNC protocol developed for the single-qubit case extends directly to arbitrary multi-qubit systems without modifying its basic structure.

As a demonstration, we consider a two-qubit probe initialized in the GHZ state,
\begin{equation}
|\mathrm{GHZ}\rangle
=\frac{|00\rangle+|11\rangle}{\sqrt2},
\end{equation}
with sensing Hamiltonian
\begin{equation}
H_s =\omega
(\sigma_z^{(1)}+\sigma_z^{(2)}).
\end{equation}
The probe accumulates Heisenberg-limited phase information during the sensing stage and subsequently enters the noisy memory channel.

\begin{figure}[htbp]
  \centering
  \includegraphics[width=1\linewidth]{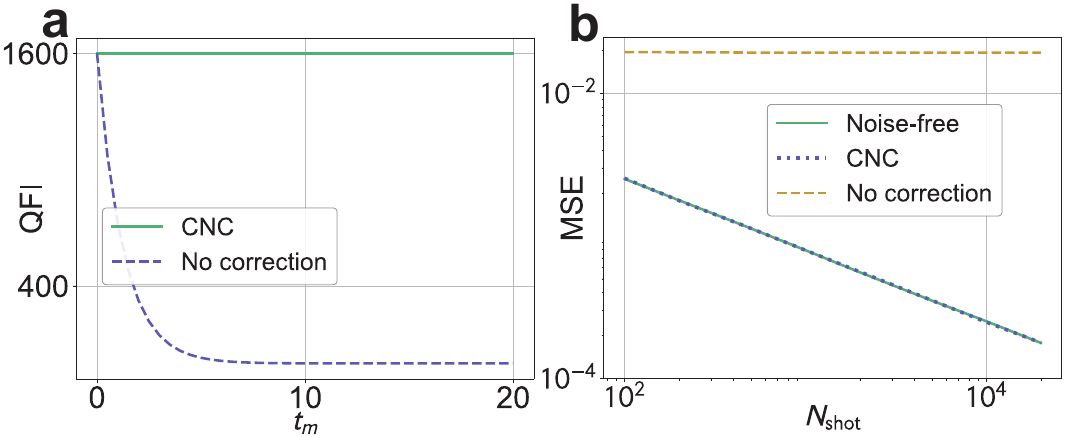}
  \caption{
  Two-qubit CNC for a single sensing-storage cycle with local dephasing noise during the memory stage.
(a) QFI as a function of memory time for a fixed sensing duration.
(b) MSE as a function of the number of measurement shots after the storage stage.
The CNC protocol is compared with the uncorrected and noise-free cases.
}
  \label{fig5}
\end{figure}

\subsection{Local Dephasing Noise}

We first consider a single sensing-storage cycle with local dephasing noise acting independently on the two probe qubits during the memory stage. The probe is first encoded by the noiseless sensing Hamiltonian and is then stored in the noisy memory channel. No further signal accumulation or quantum processing is performed after this storage stage.

To protect the stored metrological information, two auxiliary qubits are introduced, each coupled to one probe qubit through the same CNC construction used in the single-qubit dephasing case. The auxiliary qubits independently record the local dephasing syndromes, and the corresponding conditional corrections are applied only once at the end of the memory stage.

The numerical results are shown in Fig.~\ref{fig5}. Figure~\ref{fig5}(a) plots the QFI as a function of memory time for a fixed sensing duration. Without correction, local dephasing suppresses the GHZ coherence and degrades the stored metrological information. In contrast, CNC preserves the ideal QFI throughout the memory interval.

Figure~\ref{fig5}(b) shows the mean-square estimation error (MSE) as a function of the number of measurement shots $N_{\rm shot}$. The final probe state is measured after the single storage stage, and the parameter is estimated using the noiseless response function. The CNC protocol is almost indistinguishable from the noise-free case and exhibits the expected decrease of MSE with increasing shot number. By contrast, the uncorrected protocol approaches a finite error floor due to the systematic distortion caused by memory noise.

\subsection{Multi-Round Signal Processing}

The previous example assumes that the probe is measured immediately after the memory stage. In many quantum-information-processing tasks, however, sensing is only an intermediate subroutine, and the stored probe must subsequently participate in further signal accumulation or quantum computation. In this case, it is not sufficient to preserve only the QFI; the probe state itself must be restored before entering the next round.

To illustrate this situation, we consider a multi-round protocol. Each round consists of a sensing stage followed by a memory stage, with single-round durations
\begin{equation}
t_s=1,\qquad t_m=2 .
\end{equation}
During every memory stage, the two probe qubits experience independent local bit-flip noise. Since the output state of one round serves as the input state of the next, bit-flip errors must be corrected after each storage stage. Accordingly, two auxiliary qubits are used to detect and correct the local bit-flip errors.

\begin{figure}[htbp]
  \centering
  \includegraphics[width=1\linewidth]{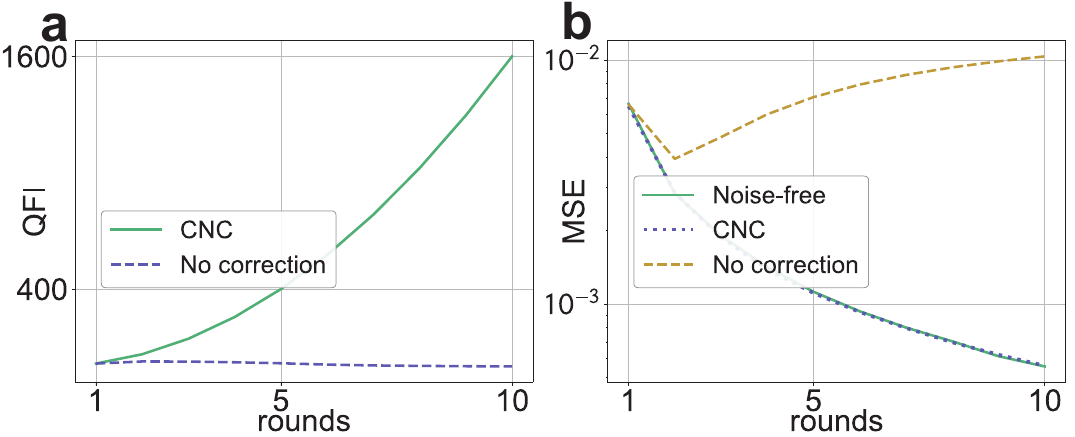}
  \caption{
  Multi-round sensing-storage protocol with local bit-flip noise during each memory stage.
Each round consists of a sensing period $t_s=1$ and a storage period $t_m=2$.
(a) QFI as a function of the number of rounds.
(b) MSE as a function of the number of rounds for $1000$ measurement shots.
}
  \label{fig6}
\end{figure}

The results are shown in Fig.~\ref{fig6}. Figure~\ref{fig6}(a) plots the QFI as a function of the number of rounds. Without correction, accumulated memory errors prevent coherent signal accumulation, and the QFI quickly deviates from the ideal quadratic growth. By contrast, CNC restores the probe after each storage stage and enables continuous accumulation of metrological information.

Figure~\ref{fig6}(b) shows the corresponding MSE for a fixed number of $1000$ measurement shots. As the number of rounds increases, the uncorrected protocol suffers from accumulated storage errors and displays a growing estimation error. The CNC protocol, on the other hand, remains almost identical to the noise-free case, confirming that the probe can be reused for repeated sensing-storage cycles without loss of metrological performance.

These results show that CNC is scalable to multi-qubit systems and naturally supports repeated sensing-storage-processing architectures, where restoring the probe state is essential for subsequent quantum operations.

\section{Conclusions}

In this work, we proposed a correlated-noise correction (CNC) protocol for protecting quantum probes during the storage stage of quantum sensing. Unlike conventional quantum error correction for quantum metrology, which suppresses noise during signal accumulation, the present protocol targets memory errors after the sensing process has already been completed. By introducing auxiliary qubits and fixed two-body entangling gates, memory errors are converted into measurable syndromes and corrected only once after the storage stage, avoiding continuous syndrome extraction throughout the memory process.
For single-qubit probes, we demonstrated that CNC effectively preserves the quantum Fisher information under dephasing, bit-flip, and amplitude-damping noise. A distinctive feature of storage-stage protection is that preserving the encoded metrological information does not necessarily require restoring the entire quantum state. When the probe is measured immediately after storage, correcting only the error components that degrade the encoded parameter is sufficient to recover the ideal sensing precision. However, when the stored probe is reused for subsequent signal accumulation or quantum-information-processing tasks, faithful recovery of the probe state becomes essential, requiring complete correction of all relevant error channels.
We further generalized the CNC protocol to multi-qubit probes and demonstrated its effectiveness for both single sensing-storage cycles and multi-round sensing-storage-processing protocols. The numerical results show that CNC preserves both the quantum Fisher information and the estimation accuracy during storage, enabling quantum probes to be repeatedly reused without degradation of metrological performance.

Our work extends quantum error correction in quantum metrology from the sensing stage to the storage stage and establishes a practical framework for protecting metrological information in quantum memories. We expect the proposed protocol to provide a useful building block for quantum-information-processing architectures involving intermediate storage of quantum probes, including iterative quantum sensing, quantum machine learning, variational quantum algorithms, and distributed quantum sensing.

\begin{acknowledgments}
The authors acknowledge support from the National Key R\&D Program of China (No.~2025YFF0515504), the National Natural Science Foundation of China (Nos.~62401359, 62471289, 62471001), the State Key Laboratory of Photonics and Communications, the Quantum Science and Technology—National Science and Technology Major Project (No.~2021ZD0300703), and the Shanghai Municipal Science and Technology Major Project (No.~2019SHZDZX01).
\end{acknowledgments}






\nocite{*}
\bibliography{sample}

@article{sensing1,
  title={Quantum metrology},
  author={Giovannetti, Vittorio and Lloyd, Seth and Maccone, Lorenzo},
  journal={Physical Review Letters},
  volume={96},
  number={1},
  pages={010401},
  year={2006},
  publisher={APS}
}

@article{sensing2,
  title={Quantum metrology with entangled coherent states},
  author={Joo, Jaewoo and Munro, William J and Spiller, Timothy P},
  journal={Physical Review Letters},
  volume={107},
  number={8},
  pages={083601},
  year={2011},
  publisher={APS}
}

@article{sensing3,
  title={Quantum sensing},
  author={Degen, Christian L and Reinhard, Friedemann and Cappellaro, Paola},
  journal={Reviews of Modern Physics},
  volume={89},
  number={3},
  pages={035002},
  year={2017},
  publisher={APS}
}

@article{sensing4,
  title={Quantum sensing of control errors in three-level systems by coherent control techniques},
  author={Xu, Hang and Song, Xue-Ke and Wang, Dong and Ye, Liu},
  journal={Science China Physics, Mechanics \& Astronomy},
  volume={66},
  number={4},
  pages={240314},
  year={2023},
  publisher={Springer}
}

@article{HL1,
  title={Toward Heisenberg Limit without Critical Slowing Down via Quantum Reinforcement Learning},
  author={Xu, Hang and Xiao, Tailong and Huang, Jingzheng and He, Ming and Fan, Jianping and Zeng, Guihua},
  journal={Physical Review Letters},
  volume={134},
  number={12},
  pages={120803},
  year={2025},
  publisher={APS}
}

@article{ico_cvsensing1,
  title={Quantum metrology with indefinite causal order},
  author={Zhao, Xiaobin and Yang, Yuxiang and Chiribella, Giulio},
  journal={Physical Review Letters},
  volume={124},
  number={19},
  pages={190503},
  year={2020},
  publisher={APS}
}

@article{ico_cvsensing2,
  title={Experimental super-Heisenberg quantum metrology with indefinite gate order},
  author={Yin, Peng and Zhao, Xiaobin and Yang, Yuxiang and Guo, Yu and Zhang, Wen-Hao and Li, Gong-Chu and Han, Yong-Jian and Liu, Bi-Heng and Xu, Jin-Shi and Chiribella, Giulio and others},
  journal={Nature Physics},
  volume={19},
  number={8},
  pages={1122--1127},
  year={2023},
  publisher={Nature Publishing Group UK London}
}

@article{nonh1,
  title={Toward Heisenberg scaling in non-Hermitian metrology at the quantum regime},
  author={Yu, Xinglei and Zhao, Xinzhi and Li, Liangsheng and Hu, Xiao-Min and Duan, Xiangmei and Yuan, Haidong and Zhang, Chengjie},
  journal={Science Advances},
  volume={10},
  number={19},
  pages={eadk7616},
  year={2024},
  publisher={American Association for the Advancement of Science}
}

@article{nonh2,
  title={Exponentially-enhanced quantum sensing with non-Hermitian lattice dynamics},
  author={McDonald, Alexander and Clerk, Aashish A},
  journal={Nature communications},
  volume={11},
  number={1},
  pages={5382},
  year={2020},
  publisher={Nature Publishing Group UK London}
}

@article{nonh3,
  title={Fundamental sensitivity limits for non-Hermitian quantum sensors},
  author={Ding, Wenkui and Wang, Xiaoguang and Chen, Shu},
  journal={Physical Review Letters},
  volume={131},
  number={16},
  pages={160801},
  year={2023},
  publisher={APS}
}

@article{phase1,
  title={Global sensing and its impact for quantum many-body probes with criticality},
  author={Montenegro, Victor and Mishra, Utkarsh and Bayat, Abolfazl},
  journal={Physical Review Letters},
  volume={126},
  number={20},
  pages={200501},
  year={2021},
  publisher={APS}
}

@article{phase2,
  title={Driving enhanced quantum sensing in partially accessible many-body systems},
  author={Mishra, Utkarsh and Bayat, Abolfazl},
  journal={Physical Review Letters},
  volume={127},
  number={8},
  pages={080504},
  year={2021},
  publisher={APS}
}

@article{noisysensing0,
  title={General framework for estimating the ultimate precision limit in noisy quantum-enhanced metrology},
  author={Escher, BM and de Matos Filho, Ruynet Lima and Davidovich, Luiz},
  journal={Nature Physics},
  volume={7},
  number={5},
  pages={406--411},
  year={2011},
  publisher={Nature Publishing Group UK London}
}

@article{noisysensing1,
  title={Entanglement-enhanced sensing in a lossy and noisy environment},
  author={Zhang, Zheshen and Mouradian, Sara and Wong, Franco NC and Shapiro, Jeffrey H},
  journal={Physical Review Letters},
  volume={114},
  number={11},
  pages={110506},
  year={2015},
  publisher={APS}
}

@article{noisysensing2,
  title={Experimental proof of quantum Zeno-assisted noise sensing},
  author={Do, Hoang-Van and Lovecchio, Cosimo and Mastroserio, Ivana and Fabbri, Nicole and Cataliotti, Francesco S and Gherardini, Stefano and M{\"u}ller, Matthias M and Dalla Pozza, Nicola and Caruso, Filippo},
  journal={New Journal of Physics},
  volume={21},
  number={11},
  pages={113056},
  year={2019},
  publisher={IOP Publishing}
}

@article{noisysensing3,
  title={Overcoming detection loss and noise in squeezing-based optical sensing},
  author={Frascella, Gaetano and Agne, Sascha and Khalili, Farid Ya and Chekhova, Maria V},
  journal={npj Quantum Information},
  volume={7},
  number={1},
  pages={72},
  year={2021},
  publisher={Nature Publishing Group UK London}
}

@article{noisysensing4,
  title={Learning to Restore Heisenberg Limit in Noisy Quantum Sensing via Quantum Digital Twin},
  author={Xu, Hang and Xiao, Tailong and Huang, Jingzheng and Fan, Jianping and Zeng, Guihua},
  journal={arXiv preprint arXiv:2508.11198},
  year={2025}
}

@article{nogo,
  title={Floquet engineering to overcome no-go theorem of noisy quantum metrology},
  author={Bai, Si-Yuan and An, Jun-Hong},
  journal={Physical Review Letters},
  volume={131},
  number={5},
  pages={050801},
  year={2023},
  publisher={APS}
}

@article{control1,
  title={Robust coherent control in three-level quantum systems using composite pulses},
  author={Xu, Hang and Song, Xue-Ke and Wang, Dong and Ye, Liu},
  journal={Optics Express},
  volume={30},
  number={2},
  pages={3125--3137},
  year={2022},
  publisher={Optica Publishing Group}
}

@article{control2,
  title={Physically feasible three-level transitionless quantum driving with multiple Schr{\"o}dinger dynamics},
  author={Song, Xue-Ke and Ai, Qing and Qiu, Jing and Deng, Fu-Guo},
  journal={Physical Review A},
  volume={93},
  number={5},
  pages={052324},
  year={2016},
  publisher={APS}
}

@article{control3,
  title={Parameter estimation in quantum sensing based on deep reinforcement learning},
  author={Xiao, Tailong and Fan, Jianping and Zeng, Guihua},
  journal={npj Quantum Information},
  volume={8},
  number={1},
  pages={2},
  year={2022},
  publisher={Nature Publishing Group UK London}
}

@article{control4,
  title={Variational principle for optimal quantum controls in quantum metrology},
  author={Yang, Jing and Pang, Shengshi and Chen, Zekai and Jordan, Andrew N and Del Campo, Adolfo},
  journal={Physical Review Letters},
  volume={128},
  number={16},
  pages={160505},
  year={2022},
  publisher={APS}
}

@article{qem1,
  title={Error-mitigated quantum metrology via virtual purification},
  author={Yamamoto, Kaoru and Endo, Suguru and Hakoshima, Hideaki and Matsuzaki, Yuichiro and Tokunaga, Yuuki},
  journal={Physical Review Letters},
  volume={129},
  number={25},
  pages={250503},
  year={2022},
  publisher={APS}
}

@article{qem2,
  title={Quantum error mitigation},
  author={Cai, Zhenyu and Babbush, Ryan and Benjamin, Simon C and Endo, Suguru and Huggins, William J and Li, Ying and McClean, Jarrod R and O’Brien, Thomas E},
  journal={Reviews of Modern Physics},
  volume={95},
  number={4},
  pages={045005},
  year={2023},
  publisher={APS}
}

@article{qem3,
  title={Machine learning for practical quantum error mitigation},
  author={Liao, Haoran and Wang, Derek S and Sitdikov, Iskandar and Salcedo, Ciro and Seif, Alireza and Minev, Zlatko K},
  journal={Nature Machine Intelligence},
  volume={6},
  number={12},
  pages={1478--1486},
  year={2024},
  publisher={Nature Publishing Group UK London}
}

@article{dd1,
  title={Dynamical decoupling leads to improved scaling in noisy quantum metrology},
  author={Sekatski, Pavel and Skotiniotis, Michalis and D{\"u}r, Wolfgang},
  journal={New Journal of Physics},
  volume={18},
  number={7},
  pages={073034},
  year={2016},
  publisher={IOP Publishing}
}

@article{dd2,
  title={Enhanced solid-state multispin metrology using dynamical decoupling},
  author={Pham, Linh My and Bar-Gill, Nir and Belthangady, Chinmay and Le Sage, David and Cappellaro, Paola and Lukin, Mikhail D and Yacoby, Amir and Walsworth, Ronald L},
  journal={Physical Review B—Condensed Matter and Materials Physics},
  volume={86},
  number={4},
  pages={045214},
  year={2012},
  publisher={APS}
}

@article{qec1,
  title={Quantum error correction for metrology},
  author={Kessler, Eric M and Lovchinsky, Igor and Sushkov, Alexander O and Lukin, Mikhail D},
  journal={Physical Review Letters},
  volume={112},
  number={15},
  pages={150802},
  year={2014},
  publisher={APS}
}

@article{qec2,
  title={Improved quantum metrology using quantum error correction},
  author={D{\"u}r, Wolfgang and Skotiniotis, Michalis and Froewis, Florian and Kraus, Barbara},
  journal={Physical Review Letters},
  volume={112},
  number={8},
  pages={080801},
  year={2014},
  publisher={APS}
}

@article{qec3,
  title={Quantum metrology enhanced by repetitive quantum error correction},
  author={Unden, Thomas and Balasubramanian, Priya and Louzon, Daniel and Vinkler, Yuval and Plenio, Martin B and Markham, Matthew and Twitchen, Daniel and Stacey, Alastair and Lovchinsky, Igor and Sushkov, Alexander O and others},
  journal={Physical Review Letters},
  volume={116},
  number={23},
  pages={230502},
  year={2016},
  publisher={APS}
}

@article{qec4,
  title={Dissipative quantum error correction and application to quantum sensing with trapped ions},
  author={Reiter, Florentin and S{\o}rensen, Anders S{\o}ndberg and Zoller, Peter and Muschik, CA},
  journal={Nature Communications},
  volume={8},
  number={1},
  pages={1822},
  year={2017},
  publisher={Nature Publishing Group UK London}
}

@article{qec5,
  title={Achieving the Heisenberg limit in quantum metrology using quantum error correction},
  author={Zhou, Sisi and Zhang, Mengzhen and Preskill, John and Jiang, Liang},
  journal={Nature Communications},
  volume={9},
  number={1},
  pages={78},
  year={2018},
  publisher={Nature Publishing Group UK London}
}

@article{qec6,
  title={Spatial noise filtering through error correction for quantum sensing},
  author={Layden, David and Cappellaro, Paola},
  journal={npj Quantum Information},
  volume={4},
  number={1},
  pages={30},
  year={2018},
  publisher={Nature Publishing Group UK London}
}

@article{qec7,
  title={Noise-resilient heisenberg-limited quantum sensing via indefinite-causal-order error correction},
  author={Xu, Hang and Deng, Xiaoyang and Zheng, Ze and Xiao, Tailong and Zeng, Guihua},
  journal={Physical Review Letters},
  volume={137},
  number={2},
  pages={020801},
  year={2026},
  publisher={APS}
}

@article{correction1,
  title={Real-time quantum error correction beyond break-even},
  author={Sivak, Volodymyr V and Eickbusch, Alec and Royer, Baptiste and Singh, Shraddha and Tsioutsios, Ioannis and Ganjam, Suhas and Miano, Alessandro and Brock, BL and Ding, AZ and Frunzio, Luigi and others},
  journal={Nature},
  volume={616},
  number={7955},
  pages={50--55},
  year={2023},
  publisher={Nature Publishing Group UK London}
}

@article{correction2,
  title={New class of quantum error-correcting codes for a bosonic mode},
  author={Michael, Marios H and Silveri, Matti and Brierley, RT and Albert, Victor V and Salmilehto, Juha and Jiang, Liang and Girvin, Steven M},
  journal={Physical Review X},
  volume={6},
  number={3},
  pages={031006},
  year={2016},
  publisher={APS}
}

@article{nisq1,
  title={Noisy intermediate-scale quantum algorithms},
  author={Bharti, Kishor and Cervera-Lierta, Alba and Kyaw, Thi Ha and Haug, Tobias and Alperin-Lea, Sumner and Anand, Abhinav and Degroote, Matthias and Heimonen, Hermanni and Kottmann, Jakob S and Menke, Tim and others},
  journal={Reviews of Modern Physics},
  volume={94},
  number={1},
  pages={015004},
  year={2022},
  publisher={APS}
}

@article{qml1,
  title={Quantum machine learning},
  author={Biamonte, Jacob and Wittek, Peter and Pancotti, Nicola and Rebentrost, Patrick and Wiebe, Nathan and Lloyd, Seth},
  journal={Nature},
  volume={549},
  number={7671},
  pages={195--202},
  year={2017},
  publisher={Nature Publishing Group UK London}
}

@article{qml2,
  title={Power of data in quantum machine learning},
  author={Huang, Hsin-Yuan and Broughton, Michael and Mohseni, Masoud and Babbush, Ryan and Boixo, Sergio and Neven, Hartmut and McClean, Jarrod R},
  journal={Nature Communications},
  volume={12},
  number={1},
  pages={2631},
  year={2021},
  publisher={Nature Publishing Group UK London}
}

@article{qml3,
  title={Challenges and opportunities in quantum machine learning},
  author={Cerezo, Marco and Verdon, Guillaume and Huang, Hsin-Yuan and Cincio, Lukasz and Coles, Patrick J},
  journal={Nature Computational Science},
  volume={2},
  number={9},
  pages={567--576},
  year={2022},
  publisher={Nature Publishing Group US New York}
}

@article{qml4,
  title={Practical advantage of quantum machine learning in ghost imaging},
  author={Xiao, Tailong and Zhai, Xinliang and Wu, Xiaoyan and Fan, Jianping and Zeng, Guihua},
  journal={Communications Physics},
  volume={6},
  number={1},
  pages={171},
  year={2023},
  publisher={Nature Publishing Group UK London}
}

@article{qml5,
  title={Quantum deep generative prior with programmable quantum circuits},
  author={Xiao, Tailong and Zhai, Xinliang and Huang, Jingzheng and Fan, Jianping and Zeng, Guihua},
  journal={Communications Physics},
  volume={7},
  number={1},
  pages={276},
  year={2024},
  publisher={Nature Publishing Group UK London}
}

@article{vqa1,
  title={Variational quantum algorithms},
  author={Cerezo, Marco and Arrasmith, Andrew and Babbush, Ryan and Benjamin, Simon C and Endo, Suguru and Fujii, Keisuke and McClean, Jarrod R and Mitarai, Kosuke and Yuan, Xiao and Cincio, Lukasz and others},
  journal={Nature Reviews Physics},
  volume={3},
  number={9},
  pages={625--644},
  year={2021},
  publisher={Nature Publishing Group UK London}
}

@article{vqa2,
  title={Variational quantum algorithms for nonlinear problems},
  author={Lubasch, Michael and Joo, Jaewoo and Moinier, Pierre and Kiffner, Martin and Jaksch, Dieter},
  journal={Physical Review A},
  volume={101},
  number={1},
  pages={010301},
  year={2020},
  publisher={APS}
}

@article{vqa3,
  title={Optimizing variational quantum algorithms using pontryagin’s minimum principle},
  author={Yang, Zhi-Cheng and Rahmani, Armin and Shabani, Alireza and Neven, Hartmut and Chamon, Claudio},
  journal={Physical Review X},
  volume={7},
  number={2},
  pages={021027},
  year={2017},
  publisher={APS}
}

@article{dqs1,
  title={Distributed quantum sensing in a continuous-variable entangled network},
  author={Guo, Xueshi and Breum, Casper R and Borregaard, Johannes and Izumi, Shuro and Larsen, Mikkel V and Gehring, Tobias and Christandl, Matthias and Neergaard-Nielsen, Jonas S and Andersen, Ulrik L},
  journal={Nature Physics},
  volume={16},
  number={3},
  pages={281--284},
  year={2020},
  publisher={Nature Publishing Group UK London}
}

@article{dqs2,
  title={Private and robust states for distributed quantum sensing},
  author={Bugalho, Lu{\'\i}s and Hassani, Majid and Omar, Yasser and Markham, Damian},
  journal={Quantum},
  volume={9},
  pages={1596},
  year={2025},
  publisher={Verein zur F{\"o}rderung des Open Access Publizierens in den Quantenwissenschaften}
}

@article{qn1,
  title={Realization of a multinode quantum network of remote solid-state qubits},
  author={Pompili, Matteo and Hermans, Sophie LN and Baier, Simon and Beukers, Hans KC and Humphreys, Peter C and Schouten, Raymond N and Vermeulen, Raymond FL and Tiggelman, Marijn J and dos Santos Martins, Laura and Dirkse, Bas and others},
  journal={Science},
  volume={372},
  number={6539},
  pages={259--264},
  year={2021},
  publisher={American Association for the Advancement of Science}
}

@article{qn2,
  title={Satellite-relayed intercontinental quantum network},
  author={Liao, Sheng-Kai and Cai, Wen-Qi and Handsteiner, Johannes and Liu, Bo and Yin, Juan and Zhang, Liang and Rauch, Dominik and Fink, Matthias and Ren, Ji-Gang and Liu, Wei-Yue and others},
  journal={Physical Review Letters},
  volume={120},
  number={3},
  pages={030501},
  year={2018},
  publisher={APS}
}

@article{crb1,
  title={Statistical distance and the geometry of quantum states},
  author={Braunstein, Samuel L and Caves, Carlton M},
  journal={Physical Review Letters},
  volume={72},
  number={22},
  pages={3439},
  year={1994},
  publisher={APS}
}

@article{crb2,
  title={Quantum Fisher information matrix and multiparameter estimation},
  author={Liu, Jing and Yuan, Haidong and Lu, Xiao-Ming and Wang, Xiaoguang},
  journal={Journal of Physics A: Mathematical and Theoretical},
  volume={53},
  number={2},
  pages={023001},
  year={2020},
  publisher={IOP Publishing}
}

\end{document}